  \documentclass[structabstract]{aa}
\usepackage{graphicx}
\usepackage{txfonts}
\usepackage{natbib}
\usepackage{amsmath}

\bibpunct[]{(}{)}{;}{a}{}{,}

\begin{document}

   \title{Rotational spectroscopy of isotopic cyclopropenone, $c$-H$_2$C$_3$O, 
          and determination of its equilibrium structure\thanks{Transition 
          frequencies from this work as well as related data from earlier work 
          are given for each isotopic species as supplementary material. 
          We  also provide quantum numbers, uncertainties, and residuals between 
          measured frequencies and those calculated from the final sets of 
          spectroscopic parameters. The data are available at CDS via anonymous 
          ftp to cdsarc.u-strasbg.fr (130.79.128.5) or via 
          http://cdsweb.u-strasbg.fr/cgi-bin/qcat?J/A+A/}}

   \author{Holger S.~P. M{\"u}ller\inst{1}
           \and
           Ananya Brahmi~M.\inst{1}
           \and
           Jean-Claude Guillemin\inst{2}
           \and
           Frank Lewen\inst{1}
           \and
           Stephan Schlemmer\inst{1}
           }

   \institute{I.~Physikalisches Institut, Universit{\"a}t zu K{\"o}ln,
              Z{\"u}lpicher Str. 77, 50937 K{\"o}ln, Germany\\
              \email{hspm@ph1.uni-koeln.de}
              \and
              Univ Rennes, Ecole Nationale Sup{\'e}rieure de Chimie de Rennes, 
              CNRS, ISCR$-$UMR 6226, 35000 Rennes, France
              }

   \date{Received 08 Dec 2020 / Accepted 21 Jan 2021}
 
  \abstract
{Cyclopropenone was first detected in the cold and less dense envelope of the giant molecular 
cloud Sagittarius~B2(N). It was found later in several cold dark clouds and it may be possible 
to detect its minor isotopic species in these environments. In addition, the main species 
may well be identified in warmer environments.}
{We aim to extend existing line lists of isotopologs of $c$-H$_2$C$_3$O from the microwave 
to the millimeter region and create one for the singly deuterated isotopolog to facilitate 
their detections in space. Furthermore, we aim to extend the line list of the main isotopic 
species to the submillimeter region and to evaluate an equilibrium structure of the molecule.}
{We employed a cyclopropenone sample in natural isotopic composition to investigate the rotational 
spectra of the main and $^{18}$O-containing isotopologs as well as the two isotopomers containing 
one $^{13}$C atom. Spectral recordings of the singly and doubly deuterated isotopic species were 
obtained using a cyclopropenone sample highly enriched in deuterium. We recorded rotational 
transitions in the 70$-$126~GHz and 160$-$245~GHz regions for all isotopologs and  also in the 
342$-$505~GHz range for the main species. Quantum-chemical calculations were carried out 
to evaluate initial spectroscopic parameters and the differences between ground-state and 
equilibrium rotational parameters in order to derive semi-empirical equilibrium structural parameters.}
{We determined new or improved spectroscopic parameters for six isotopologs and structural 
parameters according to different structure models.}
{The spectroscopic parameters are accurate enough to identify minor isotopic species at 
centimeter and millimeter wavelengths while those of the main species are deemed to be reliable 
up to 1~THz. Our structural parameters differ from earlier ones. The deviations are attributed 
to misassignments in the earlier spectrum of one isotopic species.}

\authorrunning{Holger S.~P. M{\"u}ller et al.}
\titlerunning{Rotational spectroscopy of isotopic $c$-H$_2$C$_3$O}

\maketitle
\hyphenation{For-schungs-ge-mein-schaft}

\section{Introduction}
\label{intro}

Cyclopropenone is one of the smallest aromatic molecules 
\citep{c-H2C3O_aromatic_1983,aromatic_three-rings_ai_2011}, after cyclopropenylidene 
($c$-C$_3$H$_2$), and the first cyclic molecule with a ketone functional group that 
has been detected in space. It was first found in the cold and less dense envelope of 
the giant star-forming region Sagittarius~B2(N) \citep{c-H2C3O_det_2006}. 
More recently, $c$-H$_2$C$_3$O was detected in four dark clouds or prestellar cores, 
TMC\nobreakdash-1, B1\nobreakdash-b, L483, and Lupus-1A \citep{more_c-H2C3O_2016}, and 
in the low-mass star-forming region L1527 \citep{L1527_2017_also_c-H2C3O}.

\citet{c-H2C3O_rot_1973} presented the first account on the rotational spectrum of cyclopropenone. 
These latter authors reported between 6 and 14 rotational transitions for five isotopic species recorded 
between 9~GHz and 40~GHz. The main isotopolog and the two isotopomers containing one $^{13}$C atom were 
studied in a sample in natural isotopic composition. Benson et al., employed isotopic enriched samples 
to investigate isotopologs containing $^{18}$O or two D. Additionally, they derived structural parameters 
of $c$-H$_2$C$_3$O and determined its dipole moment and rotational $g$ values through Stark and Zeeman 
measurements, respectively. Later, \citet{c-H2C3O_rot_1990} extended the data set of the main isotopic 
species to 247~GHz. These data are sufficient to identify $c$-H$_2$C$_3$O in cold astronomical sources 
up to the lower submillimeter region, possibly up to $\sim$500~GHz, but are not enough for searches 
in warmer astronomical environments. Although cyclopropenone has so far only been identified in colder 
sources, it may well be detected in the warmer and denser parts of star-forming regions in the future. 
Propanal \citep{propanal_etc_2017}, propene, and propenal \citep{propenal_propene_2020} are recent examples 
of molecules initially found only in cold astronomical sources later found in the hot corino, the warmer 
and denser parts of a low-mass star-forming region---of IRAS 16293$-$2422~B in these latter cases. 
The isotopic data may be suitable for astronomical searches in cold environments in the microwave and 
possibly lower millimeter regions if good estimates of the lowest order centrifugal distortion parameters 
are available, but are certainly too limited for searches throughout the entire millimeter region. 
Moreover, no data are available for $c$-HDC$_3$O, which is the most promising isotopolog 
to be found in cold dark molecular clouds. Therefore, we investigated the rotational spectra 
of five isotopologs of cyclopropenone in the millimeter region to facilitate their detection 
in space and extended measurements of the main isotopic species to the lower submillimeter 
region to enable more secure searches in warmer astronomical sources.

\section{Experimental details}
\label{exptl}

\subsection{Sample preparation}
\label{syntheses}

The synthesis of normal cyclopropenone was the same as in \citet{c-H2C3O_rot_1990} and 
followed the procedure described by \citet{c-H2C3O_org-synth_1977}. The synthesis of the 
cyclopropenone sample highly enriched in deuterium followed largely \citet{c-H2C3O_synth_1972} 
with \textit{n}-Bu$_3$SnH replaced by \textit{n}-Bu$_3$SnD. Please note that cyclopropenone 
decomposes at room temperature. It is very stable in a sealed container at temperatures below 
$\sim$240~K. We kept the sample in dry ice or liquid nitrogen  for our experiments.

\subsection{Spectroscopic measurements}
\label{lab-spec}

The investigation of the rotational spectra of cyclopropenone isotopologs was carried out with 
two different spectrometers. We employed two 7~m coupled glass cells, each with an inner diameter 
of  10~cm, in a double path arrangement for measurements in the 70$-$126~GHz region, 
yielding an optical path length of 28~m. We used a 5~m double path cell with a 10~cm 
inner diameter for the 160$-$245~GHz and the 342$-$505~GHz ranges. Both spectrometers use 
Virginia Diode, Inc. (VDI), frequency multipliers driven by Rohde \& Schwarz SMF~100A 
synthesizers as sources, and Schottky diode detectors. Frequency modulation was employed 
to reduce baseline effects with demodulation at twice the modulation frequency. 
This causes absorption lines to appear approximately as second derivatives of a Gaussian. 
Additional information on the spectrometers is available in \citet{n-BuCN_rot_2012} and 
\citet{OSSO_rot_2015}, respectively.

We recorded individual transition frequencies in all three frequency windows covering 10~MHz 
for well-predicted lines up to 100~MHz in the search for first lines of the singly deuterated 
isotopolog. The pressure was around 1.0~Pa in the 3~mm region as test measurements showed that 
the peak intensity was best between $\sim$0.75 and $\sim$1.5~Pa. The pressure was between 1.0 
and 2.0~Pa at shorter wavelengths.
We refilled the cells after roughly one hour because the molecule is only moderately stable 
at room temperature; its half-life was close to one hour in our cells. Uncertainties were 
evaluated mostly based on the symmetry of the line shape and were as small as 5~kHz for 
isolated and very symmetric lines. Such small uncertainties were achieved earlier, for example, 
in the case of 2-cyanobutane with a much richer rotational spectrum \citep{2-CAB_rot_2017}.
Uncertainties for good lines were 10$-$20 kHz, and larger uncertainties up to $\sim$100~kHz 
were used for example for weaker lines and lines close to other lines.

Initial measurements using the highly deuterated cyclopropenone sample yielded no clear signs 
of rotational transitions of $c$-D$_2$C$_3$O, but instead strong signals of $c$-H$_2$C$_3$O. 
We suspected rapid D-to-H exchange on the cell walls, even though this may appear unusual 
for a molecule with substantial aromatic character. On the other hand, at least some 
unsaturated molecules are known to exchange D and H quite readily; for example, 
the reaction between HC$_3$N and D$_2$O was used to generate DC$_3$N \citep{DC3N_FTMW_2008}. 
We conditioned the cell walls with $\sim$200~Pa D$_2$O for two hours and observed strong 
signals of $c$-D$_2$C$_3$O and very weak ones of $c$-H$_2$C$_3$O afterwards. The signals 
of $c$-HDC$_3$O, identified later, were in between, roughly a factor of four weaker 
than those of $c$-D$_2$C$_3$O.

\section{Quantum-chemical calculations}
\label{qcc}

We carried out quantum-chemical calculations at the Regionales Rechenzentrum der Universit{\"a}t 
zu K{\"o}ln (RRZK) using the commercially available program Gaussian~09 \citep{Gaussian09E}. 
We performed B3LYP hybrid density functional \citep{Becke_1993,LYP_1988}, M{\o}ller-Plesset 
second- (MP2) and third-order perturbation theory (MP3) calculations \citep{MPn_1934}, 
along with coupled cluster calculations with single and double excitations augmented 
by a perturbative correction for triple excitations, CCSD(T) \citep{CC+T_1989}. 
We employed correlation consistent basis sets which were augmented with diffuse basis functions 
aug-cc-pVXZ \citep[X = T, Q, 5;][]{cc-pVXZ_1989}, which we abbreviate here to 3a, 4a, and 5a, 
respectively. These basis sets were further augmented with core-correlating basis functions 
in some cases, yielding the aug-cc-pwCVXZ basis sets \citep{core-corr_2002}, which we 
denote 3aC, 4aC, and 5aC, respectively.

Equilibrium geometries were determined by analytic gradient techniques, harmonic force fields 
by analytic second derivatives, and anharmonic force fields by numerical differentiation 
of the analytically evaluated second derivatives of the energy. The main goals of these 
anharmonic force field calculations were to evaluate initial spectroscopic parameters 
for the minor isotopic species of cyclopropenone and first-order vibration-rotation parameters 
\citep{vib-rot_rev_1972}; see also Sect.~\ref{structure}. Core electrons were kept frozen 
in MP2, MP3, and CCSD(T) calculations unless ``ae'' indicates that all electrons were correlated.


\begin{figure}
\centering
\includegraphics[width=6cm,angle=0]{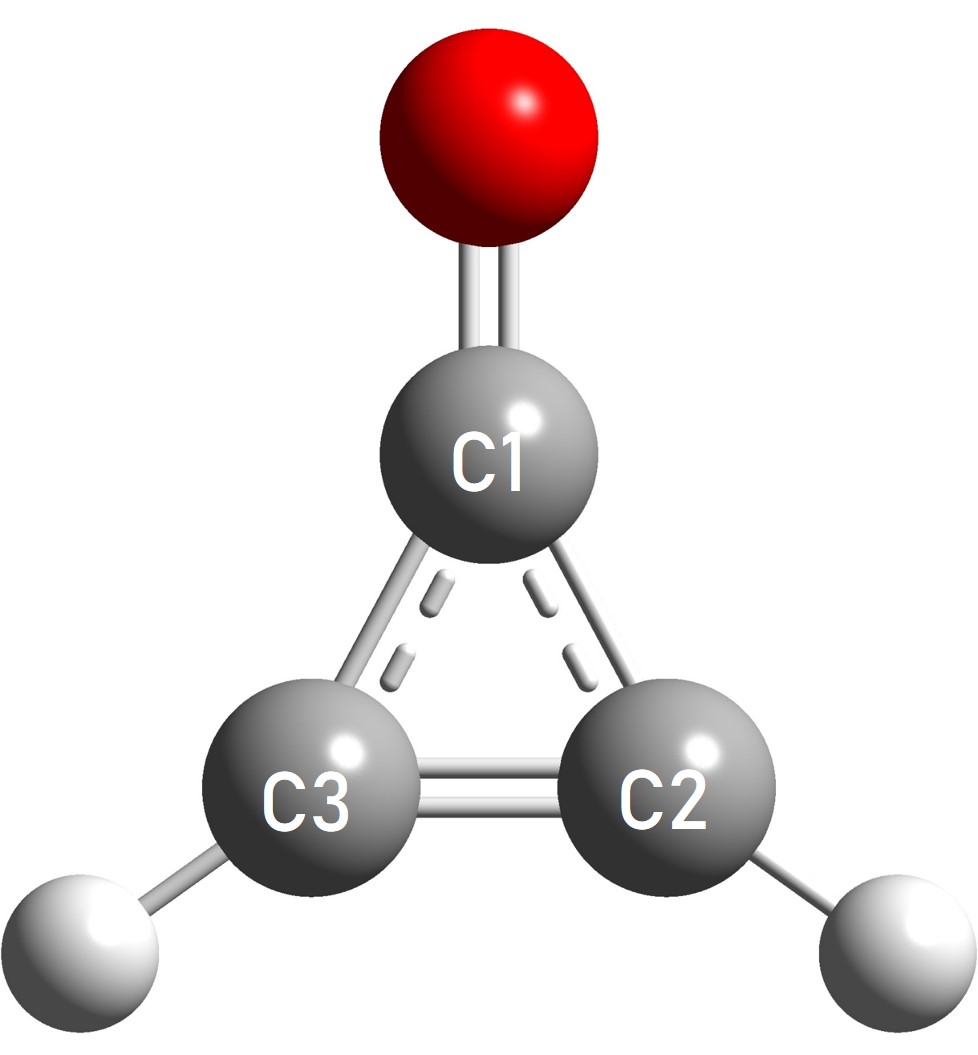}

\caption{Model of the cyclopropenone molecule. Carbon atoms are symbolized by gray spheres 
   which are numbered. Hydrogen atoms are indicated by small, light gray spheres and 
   the oxygen atom by a red sphere.}
\label{mol-bild}
\end{figure}

\section{Spectroscopic properties of cyclopropenone}
\label{rot_backgr}

Cyclopropenone is a planar asymmetric rotor with $\kappa = (2B - A - C)/(A - C) = -0.8801$, 
which is somewhat close to the prolate symmetric limit of $-$1. Its dipole moment of 4.39~D 
\citep{c-H2C3O_rot_1973} is along the $a$-inertial axis which is aligned with the CO bond; 
see Fig.~\ref{mol-bild}. The strongest transitions are $R$-branch transitions 
($\Delta J = +1$) with $\Delta K_a = 0$ and $\Delta K_c = +1$. The Boltzmann peak 
at 300~K is near 480~GHz. $Q$-branch transitions with $\Delta K_a = 0$ and 
$\Delta K_c = -1$ and transitions with $\Delta K_a = \pm2$ are much weaker, but can be 
observed quite readily under favorable circumstances.

Carbon has two stable isotopes with mass numbers 12 and 13 and with terrestrial abundances 
of 98.89\% and 1.11\%, respectively \citep{iso-comp_2011}. The respective abundances are 
99.76\%, 0.04\%, and 0.20\% for $^{16}$O, $^{17}$O, and $^{18}$O, and 99.98\% and 
$\sim$0.015\% for H and D.

\begin{figure}
\centering
\includegraphics[width=8cm,angle=0]{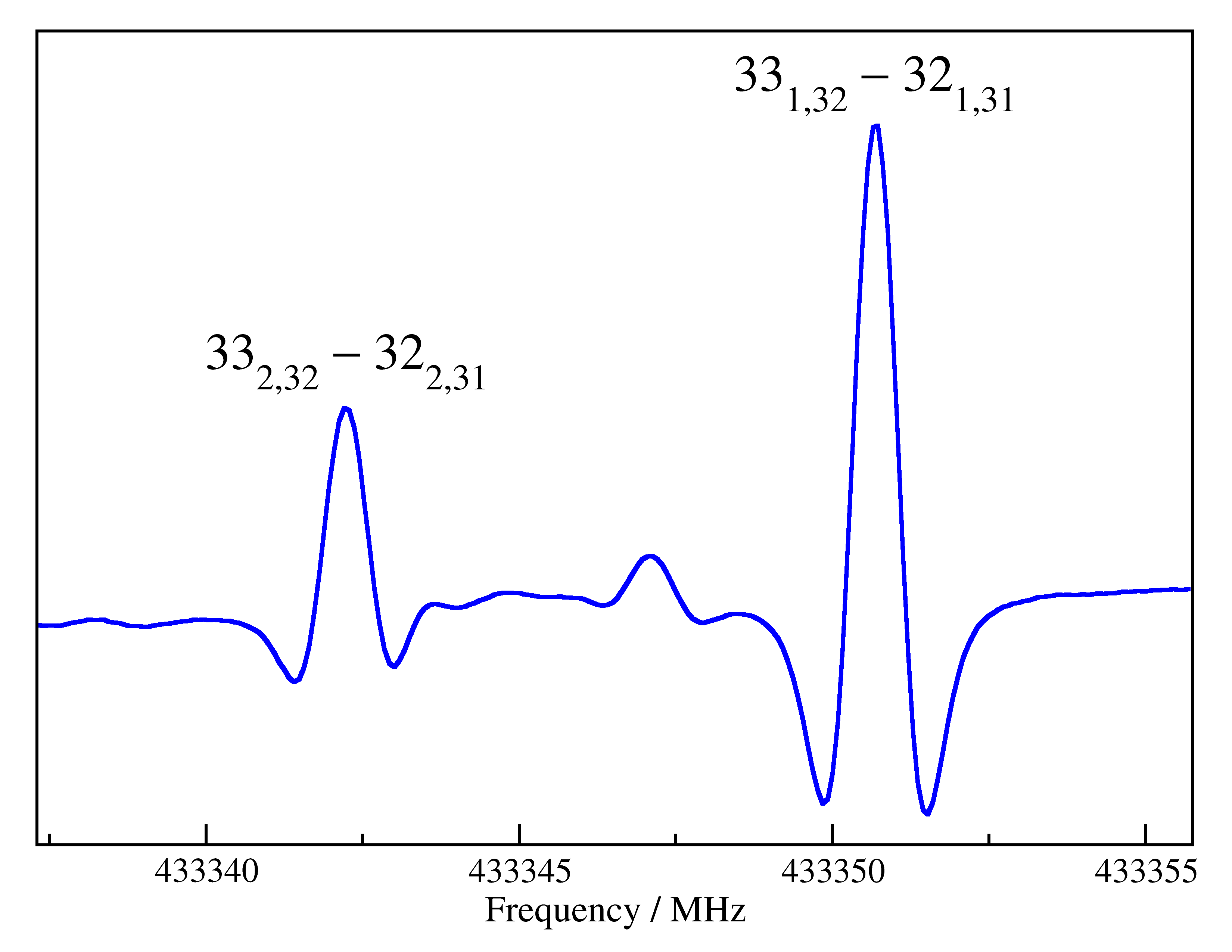}

\caption{Section of the submillimeter spectrum of $c$-H$_2$C$_3$O. The $K_c = J - 1$ transitions 
   are approaching oblate pairing, that is, the two lines coalesce at somewhat higher $J$. Also 
   demonstrated is the 1:3 \textit{para} to \textit{ortho} ratio for transitions with $K_a$ 
   being even and odd, respectively; see Sect.~\ref{rot_backgr} for further details.}
\label{spin-statistic_H2}
\end{figure}

The main isotopolog and those with $^{18}$O or $^{13}$C in the keto group (C1) have 
$C_{\rm 2v}$ symmetry. Spin-statistics caused by the two equivalent H lead to \textit{ortho} 
and \textit{para} states with a 3:1 intensity ratio; see Fig.~\ref{spin-statistic_H2}. 
The \textit{ortho} states are described by $K_a$ being odd. The doubly deuterated isotopolog also 
has $C_{\rm 2v}$ symmetry, but the \textit{ortho} to \textit{para} ratio is 2:1, 
and the \textit{ortho} states are described by $K_a$ being even. The effect of substituting 
$^{13}$C at C2 or C3 is the same because the two carbon atoms are structurally equivalent. 
The corresponding species is therefore referred to as the $^{13}$C2 isotopolog. We note that we do not 
consider isotopologs with more than one $^{13}$C. The abundance of the $^{13}$C2 isotopolog 
is $\sim$2.2\% in a sample of natural isotopic composition. Its symmetry is $C_{\rm S}$, 
as is that of the HD isotopolog. No nontrivial spin statistics exist in these two isotopologs. 
The HD isotopolog has a very small $b$-dipole moment component of $\sim$0.21~D as derived 
from the structure and from quantum-chemical calculations; the corresponding value for 
the $^{13}$C2 isotopolog is $\sim$0.06~D.


\section{Spectroscopic results}
\label{lab-results}

\begin{table*}
\begin{center}
\caption{Present and previous experimental spectroscopic parameters (MHz) of the main isotopolog of cyclopropenone 
         and details of the fits in comparison to values calculated at the B3LYP/3a level.}
\label{tab-main-species}
\renewcommand{\arraystretch}{1.10}
\begin{tabular}[t]{lr@{}lr@{}lcr@{}lr@{}lr@{}ll}
\hline \hline
 \multicolumn{5}{c}{S reduction} & & \multicolumn{7}{c}{A reduction} \\
\cline{1-5} \cline{7-13}
Parameter & \multicolumn{2}{c}{Present} & \multicolumn{2}{c}{B3LYP/3a$^a$} & & \multicolumn{2}{c}{B3LYP/3a$^a$} & \multicolumn{2}{c}{Present} 
& \multicolumn{2}{c}{Previous$^b$} & Parameter \\
\hline
$A$                  &  32040&.68588~(49)  &  32262&.52 & &  32262&.52 & 32240&.68387~(51)  & 32240&.734~(25)   & $A$                       \\
$B$                  &   7825&.000761~(53) &   7823&.00 & &   7823&.01 &  7825&.045789~(63) &  7825&.04620~(60) & $B$                       \\
$C$                  &   6280&.728112~(50) &   6287&.49 & &   6287&.48 &  6287&.684168~(62) &  6287&.68461~(57) & $C$                       \\
$D_K \times 10^3$    &     46&.4328~(91)   &     44&.85 & &     46&.47 &    48&.1139~(97)   &    50&.65~(104)   & $\Delta_K \times 10^3$    \\
$D_{JK} \times 10^3$ &     35&.87301~(90)  &     35&.27 & &     33&.33 &    33&.81184~(90)  &    33&.7882~(50)  & $\Delta_{JK} \times 10^3$ \\
$D_J \times 10^6$    &   1451&.833~(24)    &   1395&.   & &   1718&.   &  1794&.927~(83)    &  1793&.62~(100)   & $\Delta_J \times 10^6$    \\
$d_1 \times 10^6$    & $-$385&.4874~(49)   & $-$366&.1  & &     21&.37 &    22&.2410~(13)   &    22&.156~(15)   & $\delta_K \times 10^3$    \\
$d_2 \times 10^6$    & $-$171&.7446~(55)   & $-$161&.5  & &    366&.1  &   385&.546~(20)    &   385&.36~(25)    & $\delta_J \times 10^6$    \\
$H_K \times 10^9$    &       &             &  $-$33&.48 & & $-$154&.9  &      &             &      &            & $\Phi_K \times 10^9$      \\
$H_{KJ} \times 10^9$ &    363&.09~(54)     &    108&.8  & &    153&.0  &    36&.1~(22)      &      &            & $\Phi_{KJ} \times 10^9$   \\
$H_{JK} \times 10^9$ &  $-$14&.98~(39)     &  $-$10&.14 & &     75&.66 &    79&.55~(83)     &      &            & $\Phi_{JK} \times 10^9$   \\
$H_J \times 10^{12}$ &       &             & $-$226&.2  & &    319&.0  &   373&.~(37)       &      &            & $\Phi_J \times 10^{12}$   \\
$h_1 \times 10^{12}$ &       &             &     95&.67 & &   1245&.   &  1595&.~(14)       &      &            & $\phi_K \times 10^9$      \\
$h_2 \times 10^{12}$ &    276&.4~(36)      &    272&.6  & &     43&.33 &    40&.97~(85)     &      &            & $\phi_{JK} \times 10^9$   \\
$h_3 \times 10^{12}$ &    193&.0~(12)      &    151&.1  & &    246&.7  &   201&.~(11)       &      &            & $\phi_J \times 10^{12}$   \\
$J_{\rm max}$        &     48&             &       &    & &       &    &    48&             &    27&            &                           \\
$K_{a,{\rm max}}$    &     29&             &       &    & &       &    &    29&             &    12&            &                           \\
No. transitions      &    463&             &       &    & &       &    &   463&             &      &            &                           \\
No. frequencies      &    370&             &       &    & &       &    &   370&             &    59&            &                           \\
rms                  &      0&.0248        &       &    & &       &    &     0&.0243        &     0&.028        &                           \\
rms (new)$^c$        &      0&.0236        &       &    & &       &    &     0&.0229        &      &            &                           \\
rms (B)$^c$          &      0&.0293        &       &    & &       &    &     0&.0292        &      &            &                           \\
rms (G)$^c$          &      0&.0312        &       &    & &       &    &     0&.0310        &      &            &                           \\
wrms$^d$             &      0&.972         &       &    & &       &    &     0&.921         &      &            &                           \\
wrms (new)$^c$       &      0&.961         &       &    & &       &    &     0&.901         &      &            &                           \\
wrms (B)$^c$         &      0&.978         &       &    & &       &    &     0&.972         &      &            &                           \\
wrms (G)$^c$         &      1&.041         &       &    & &       &    &     1&.034         &      &            &                           \\
\hline
\end{tabular}
\end{center}
\tablefoot{
Watson's $S$ and $A$ reductions were used in the representation $I^r$. Numbers in parentheses are one standard 
deviation in units of the least significant figures. Empty fields indicate parameters not used in the fit.
$^{(a)}$ Ground-state rotational and equilibrium centrifugal distortion parameters. 
$^{(b)}$ \citet{c-H2C3O_rot_1990}.
$^{(c)}$ The labels (new), (B), and (G) refer to our new data, those from \citet{c-H2C3O_rot_1973}, and from \citet{c-H2C3O_rot_1990}. 
$^{(d)}$ Weighted rms, unit-less.         
} 
\end{table*}


Pickett's programs SPCAT and SPFIT \citep{spfit_1991} were used to calculate and fit the 
rotational spectra of the cyclopropenone isotopologs. The results of \citet{c-H2C3O_rot_1990} 
were used for the main species. Too few transition frequencies have been published for 
the minor isotopic species to reliably determine quartic centrifugal distortion parameters. 
We evaluated ground-state rotational and equilibrium quartic centrifugal distortion parameters 
for six isotopic species of cyclopropenone at the B3LYP/3a quantum-chemical level; 
vibrational corrections to centrifugal distortion parameters are, to our knowledge, 
not available in publicly available quantum-chemical programs. 
We approximated the experimental value $X_{\rm exp}^{\rm iso}$ of isotopic spectroscopic 
parameters by scaling the calculated value $X_{\rm calc}^{\rm iso}$ with the ratio of the 
experimental and calculated value $X_{\rm exp}^{\rm main}$/$X_{\rm calc}^{\rm main}$ 
of the main isotopolog. These values were used as starting parameters to reproduce 
the limited sets of transition frequencies of the minor isotopic species. 
One important aspect in this procedure, and in all of our fitting, was to determine 
or to float as few parameters as possible to improve the weighted rms (wrms) of the fit 
as a useful measure of the quality of the fit. This implies that we searched in each 
fitting round for the spectroscopic parameter that led to the greatest reduction of the wrms, 
and that was useful in the context of parameters already determined or floated in the fit. 
It was sufficient to float $B$ and $C$ in the cases of the $^{13}$C1, $^{13}$C2, and 
$^{18}$O isotopologs of cyclopropenone. One transition of the $^{18}$O species, 
$4_{1,3} - 4_{1,4}$, and two transitions of the $^{13}$C2 species, $10_{2,8} - 10_{2,9}$ 
and $15_{3,12} - 15_{3,13}$, had large residuals between the reported experimental 
frequency and the calculated one that we decreased the weights of these transitions 
initially; eventually, these were omitted from the final fits. The reported frequencies 
were larger than in our final calculations by 0.46~MHz for the line attributed to 
the $^{18}$O species and by 43.51 and 92.56~MHz, respectively, for the two lines 
assigned to the $^{13}$C2 isotopolog. We also had to float $A$ in order to 
satisfactorily reproduce the transition frequencies of the doubly deuterated isotopolog. 
The adjusted rotational parameters of $c$-D$_2$C$_3$O differed somewhat more from the 
initial parameters than the adjusted $B$ and $C$ values of the other minor isotopic species. 
We assumed the deviations in the case of $c$-HDC$_3$O would be roughly half as large 
as those of $c$-D$_2$C$_3$O. We corrected the $c$-HDC$_3$O accordingly to improve 
the calculations of this isotopolog for the first searches of rotational transitions.

\subsection{The main isotopic species}
\label{main_species}

Our investigations started with measurements of the main isotopic species in the 
342$-$505~GHz region. We targeted the strongest $R$-branch transitions with $K_a \le 5$ 
in a first step, extending to $K_a = 29$ in subsequent steps. Transitions with 
$K_a \le 14$ were found within 0.3~MHz of the initial calculations. The deviations 
increased rapidly with $K_a$ and exceeded 10~MHz for the highest $K_a$ transitions. 
We then searched for $Q$-branch transitions with $\Delta K_a = 0$ and for various types 
of $\Delta K_a = 2$ transitions, most of them had also $\Delta J = 0$. Later, very 
limited measurements were carried out in the 160$-$245~GHz region and then again 
more extensive ones in the 70$-$126~GHz region. The $\Delta K_a = \Delta J = 0$ 
transitions reached $K_a = 8$, and the $\Delta K_a = 2$ transitions extended to 
$K_a = 5 \leftrightarrow 7$. Whereas the rotational spectrum of cyclopropenone is 
quite sparse on the level of the strongest lines, it is much richer on the level 
of the weakest transitions recorded, which means  that the desired line has 
an increased chance of being blended with or being close to a usually unassigned line.

In the end, we recorded 398 different, novel  transitions, which corresponds to 311 
different frequencies because of 87 unresolved asymmetry doublets. The majority of 
these doublets are prolate paired transitions, which have the same $J$ and $K_a$ for each 
pair, as may be expected for an asymmetric rotor somewhat close to the prolate symmetric limit; 
seven are oblate paired transitions with $K_c = J$ in each case. In addition, we remeasured 
three transitions already reported by \citet{c-H2C3O_rot_1990}. These data were combined with 
14 transition frequencies from \citet{c-H2C3O_rot_1973} and 45 from \citet{c-H2C3O_rot_1990}, 
which corresponded to 51 transitions. Uncertainties of 30~kHz were used for these lines as 
stated by \citet{c-H2C3O_rot_1990} for their data and commensurate for the data of 
\citet{c-H2C3O_rot_1973}.

We subjected the transition frequencies initially to fits employing Watson's A reduction, 
as done previously \citep{c-H2C3O_rot_1990}. We also tried Watson's S reduction 
because cyclopropenone is quite close to the prolate symmetric limit, in which case the 
S reduction is usually preferable. As can be seen in Table~\ref{tab-main-species}, we 
required an almost complete parameter set up to sixth order in the A reduction; only 
$\Phi _K$ was not used in the fit. Two fewer parameters were used in the S reduction, 
albeit with a slightly larger wrms of 0.972 compared to 0.921. Nevertheless, we decided 
to view the S reduced fit as the preferred one. Trial fits with $\Phi _K$ or $H_K$, 
$H_J$, and $h_1$ from B3LYP/3a calculations added to the fit as fixed parameters did 
not improve the quality of the fits and resulted in changes of the remaining 
spectroscopic parameters roughly corresponding to their uncertainties. Moreover, we 
were uncertain as to the reliability of the quantum-chemically calculated sextic 
distortion parameters; see also Sect.~\ref{disc_parameters}. Therefore, these parameters 
were omitted from the final fits. The quantum-chemically calculated spectroscopic 
parameters are given in Table~\ref{tab-main-species} for comparison,  along with 
the parameters from \citet{c-H2C3O_rot_1990}. 
We also provide rms values for the whole data set and for the individual sources 
for completeness. The rms is meaningful if only one uncertainty was used in all 
instances or if the uncertainties differ only by a factor of a few. In the present case, 
the uncertainties differ by around a factor of ten. In this and similar cases, the rms 
is dominated by the lines with large residuals.


\begin{table}
\begin{center}
\caption{Spectroscopic parameters (MHz) from a minimum parameter set, a maximum parameter set, 
         and initial parameters along with details of the fits of the $^{13}$C1, $^{13}$C2, and 
         $^{18}$O isotopologs of cyclopropenone.}
\label{tab-rare-isos}
\renewcommand{\arraystretch}{1.06}
\begin{tabular}[t]{lr@{}lr@{}lr@{}l}
\hline \hline
Parameter & \multicolumn{2}{c}{Minimum} & \multicolumn{2}{c}{Maximum} & \multicolumn{2}{c}{Initial} \\
\hline
\multicolumn{7}{l}{$^{13}$C1 isotopolog}                                           \\
\hline
$A$                  &  32040&.141~(4)       &  32040&.176~(18)      &  32040&.226 \\
$B$                  &   7816&.5529~(2)      &   7816&.5525~(2)      &   7816&.565 \\
$C$                  &   6275&.2893~(2)      &   6275&.2894~(2)      &   6275&.289 \\
$D_K \times 10^3$    &     46&.61            &     49&.9~(26)        &     46&.61  \\
$D_{JK} \times 10^3$ &     35&.636~(6)       &     35&.634~(8)       &     35&.659 \\
$D_J \times 10^6$    &   1453&.62~(34)       &   1452&.67~(71)       &   1454&.9   \\
$d_1 \times 10^6$    & $-$385&.54            & $-$384&.74~(29)       & $-$385&.54  \\
$d_2 \times 10^6$    & $-$170&.65            & $-$170&.99~(38)       & $-$170&.65  \\
$J_{\rm max}$        &     21&~(15)          &     21&~(15)          &       &     \\
$K_{a,{\rm max}}$    &      7&~(3)           &      7&~(3)           &       &     \\
No. trans.           &    115&~(6,$-$0)$^c$  &    115&~(6,$-$0)$^a$  &       &     \\
No. freqs.           &    110&~(6,$-$0)$^c$  &    110&~(6,$-$0)$^a$  &       &     \\
rms                  &      0&.0271          &      0&.0186          &       &     \\
wrms$^b$             &      0&.782           &      0&.722           &       &     \\
\\
\multicolumn{7}{l}{$^{13}$C2 isotopolog}                                           \\
\hline
$A$                  &  31174&.389~(6)       &  31173&.391~(6)       &  31175&.000 \\
$B$                  &   7709&.4761~(1)      &   7709&.4759~(2)      &   7709&.598 \\
$C$                  &   6172&.7465~(1)      &   6172&.7466~(2)      &   6172&.943 \\
$D_K \times 10^3$    &     44&.80~(41)       &     44&.36~(47)       &     42&.79  \\
$D_{JK} \times 10^3$ &     34&.656           &     34&.654~(7)       &     34&.656 \\
$D_J \times 10^6$    &   1402&.36~(33)       &   1402&.22~(35)       &   1403&.8   \\
$d_1 \times 10^6$    & $-$376&.82            & $-$376&.55~(13)       & $-$376&.82  \\
$d_2 \times 10^6$    & $-$169&.59~(15)       & $-$169&.57~(18)       & $-$169&.63  \\
$J_{\rm max}$        &     30&~(6)           &     30&~(6)           &       &     \\
$K_{a,{\rm max}}$    &      7&~(1)           &      7&~(1)           &       &     \\
No. trans.           &    106&~(5,$-$2)$^c$  &    106&~(5,$-$2)$^a$  &       &     \\
No. freqs.           &    100&~(5,$-$2)$^c$  &    100&~(5,$-$2)$^a$  &       &     \\
rms                  &      0&.0184          &      0&.0180          &       &     \\
wrms$^b$             &      0&.796           &      0&.766           &       &     \\
\\
\multicolumn{7}{l}{$^{18}$O isotopolog}                                            \\
\hline
$A$                  &  32040&.339           &  32240&.323~(53)      &  32040&.339 \\
$B$                  &   7344&.8771~(3)      &   7344&.8759~(7)      &   7344&.714 \\
$C$                  &   5967&.5319~(3)      &   5967&.5326~(6)      &   5967&.034 \\
$D_K \times 10^3$    &     49&.85            &     49&.85            &     49&.85  \\
$D_{JK} \times 10^3$ &     32&.611           &     32&.594~(17)      &     32&.611 \\
$D_J \times 10^6$    &   1308&.6             &   1307&.9~(12)        &   1308&.6   \\
$d_1 \times 10^6$    & $-$324&.81            & $-$322&.8~(13)        & $-$324&.81  \\
$d_2 \times 10^6$    & $-$138&.10            & $-$136&.8~(10)        & $-$138&.10  \\
$J_{\rm max}$        &     19&~(14)          &     19&~(14)          &       &     \\
$K_{a,{\rm max}}$    &      5&~(3)           &      5&~(3)           &       &     \\
No. trans.           &     61&~(10,$-$1)$^c$ &     61&~(10,$-$1)$^a$ &       &     \\
No. freqs.           &     58&~(10,$-$1)$^c$ &     58&~(10,$-$1)$^a$ &       &     \\
rms                  &      0&.0545          &      0&.0525          &       &     \\
wrms$^b$             &      0&.871           &      0&.801           &       &     \\
\hline
\end{tabular}
\end{center}
\tablefoot{
Watson's $S$ reduction was used in the representation $I^r$. Numbers in parentheses of parameters are 
one standard deviation in units of the least significant figures. Numbers in parentheses associated with 
numbers of quantum numbers, transitions, and lines, respectively, refer to data from the previous study 
\citep{c-H2C3O_rot_1973}. Parameters without uncertainties were estimated and kept fixed in the analyses; 
see Sect.~\ref{lab-results}.
$^{(a)}$ With omitted lines after the comma; see Sect.~\ref{lab-results}.
$^{(b)}$ Weighted rms, unitless. 
}
\end{table}

\subsection{The $^{13}$C1, $^{13}$C2, and $^{18}$O isotopologs}
\label{13C_18O}

We began our investigations of the minor isotopologs of cyclopropenone containing one $^{13}$C 
or $^{18}$O in the 160$-$245~GHz region searching for the stronger $R$-branch transitions first 
with $K_a \le 2$ and then up to $K_a = 5$. Additional measurements were carried out in the 
70$-$126~GHz region, in which we recorded $R$-branch transitions up to $K_a = 7$ and several 
$\Delta K_a = \Delta J = 0$ transitions for the two isotopomers with one $^{13}$C. 
The extent of measured lines was more limited for the $^{18}$O because of its lower 
abundance.

The final line lists of the $^{13}$C1, $^{13}$C2, and $^{18}$O isotopologs consist of 
104, 95, and 48 different and new transition frequencies, respectively, in addition to 
the small number of previously reported lines \citep{c-H2C3O_rot_1973} for which 
uncertainties of 70~kHz were assigned. The resulting sets of spectroscopic parameters, 
determined as described at the beginning of this section, had not only $B$ and $C$ 
floated for $^{13}$C1 and $^{13}$C2, but also $A$ and two respectively three quartic 
distortion parameters. It was sufficient to float only $B$ and $C$ in the case of 
the $^{18}$O isotopolog. The parameter values, their uncertainties, and additional 
details of the fits are presented in Table~\ref{tab-rare-isos} as minimum parameter 
sets which are the preferred parameter sets. We also tested the number of parameters 
that can be determined with sufficient significance. These were all quartic centrifugal 
distortion parameters for the isotopomers with one $^{13}$C and all but $\Delta D_K$ 
for the $^{18}$O isotopolog. The resulting values are given as maximum parameter sets 
in the same table along with the initial parameters whose derivation was also described 
at the beginning of this section.


\begin{figure}
\centering
\includegraphics[width=9cm,angle=0]{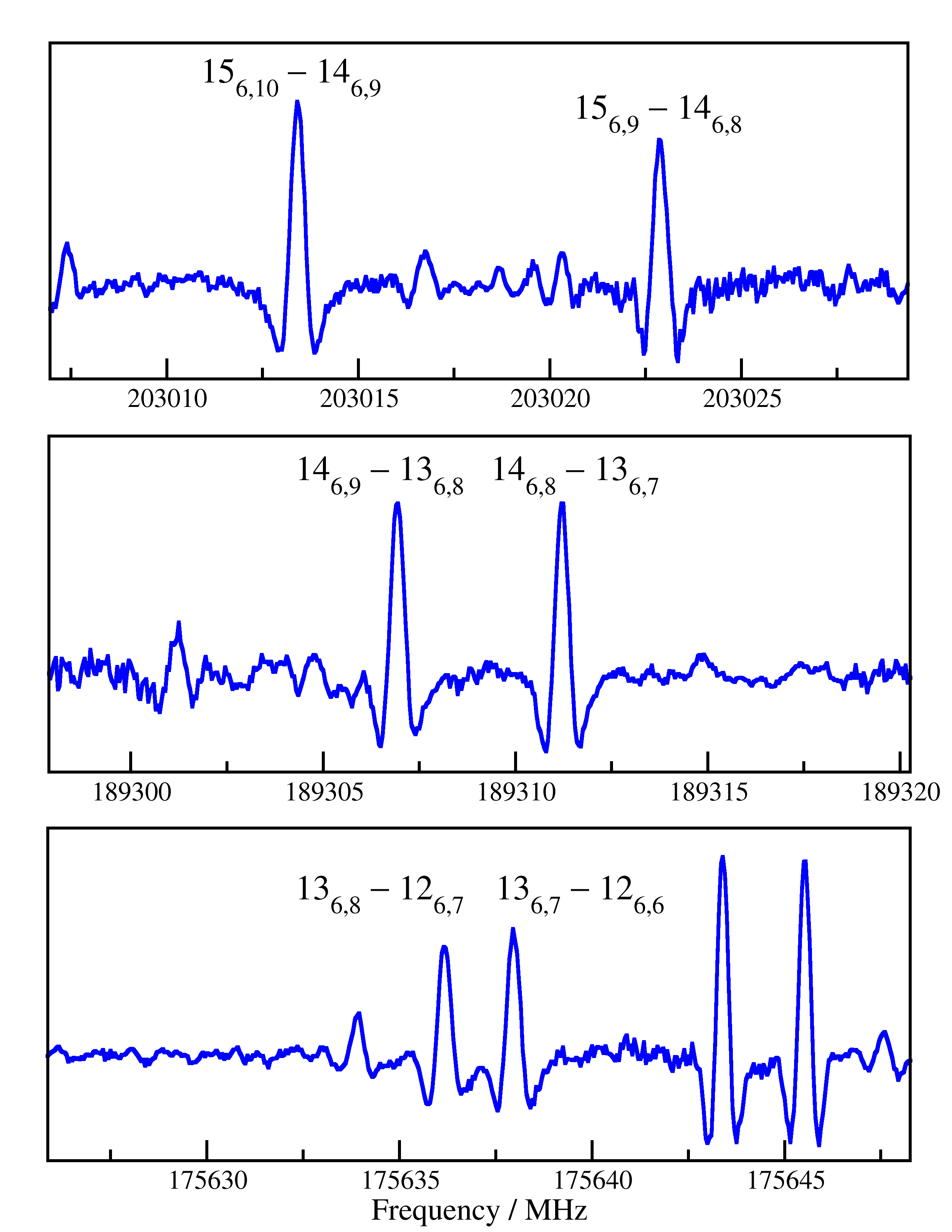}

\caption{Section of the millimeter spectrum of cyclopropenone highly enriched in deuterium. 
   Transitions with small but rapidly increasing asymmetry splitting are shown for $K_a = 6$ 
   of $c$-HDC$_3$O. We assigned these transitions first because of their 
   easily recognizable patterns.}
\label{HD-species_K-6}
\end{figure}

\subsection{The singly and doubly deuterated isotopic species}
\label{deuterated}

Our procedure for studying the rotational spectrum of the D$_2$ isotopolog was very similar to 
that for the minor isotopic species described in the previous section. The use of a sample highly 
enriched in deuterium made it possible to record considerably more transitions extending to 
higher quantum numbers. The spectral recordings were richer in lines than those of the sample in 
natural isotopic composition because of the presence of lines caused by the HD species and 
even some of the main species. In addition, we identified lines of traces of CH$_2$Cl$_2$ 
which was used as a solvent in the course of the preparation of the deuterated sample and 
was difficult to remove completely.

The $R$-branch transitions with $\Delta K_a = 0$ were recorded up to $K_a = 9$, transitions with 
$\Delta K_a = \Delta J = 0$ up to 2, and those with $\Delta K_a = 2$ up to $K_a = 5 \leftarrow 3$.

We employed a different strategy to search for transitions of the HD isotopolog because no 
previous data exist. A pattern of two or more transitions is obviously more decisive than 
a single line. The best candidates are nearly prolate or nearly oblate paired $R$-branch 
transitions because the small asymmetry splitting changes rapidly with $J$. The asymmetry 
splitting of the $K_c = J$ nearly oblate paired transitions was deemed to be too large 
in the 160$-$245~GHz region. As can be seen in Fig.~\ref{HD-species_K-6}, the $K_a = 6$ 
nearly prolate paired transitions were suited very well, even though they were considerably 
weaker. The measured asymmetry splittings of about 1.8, 4.4, and 9.5~MHz for $J" = 12$ to 14 
compared very well with the calculated splittings. We searched for $R$-branch transitions 
with equal or lower $K_a$ with an improved calculation of the rotational spectrum. 
Subsequently, we sought $Q$-branch transitions with $\Delta K_a = 0$ and 2, respectively. 
Our line list reached $K_a = 3$ in the first case and $K_a = 3 \leftarrow 1$ in the second 
case of these weaker transitions.

We did not target any $b$-type transitions of $c$-HDC$_3$O as these were calculated to be 
even weaker than many of the $\Delta K_a = 2$ transitions. Some $b$-type transitions were 
covered by accident; even much longer integration times may not have helped because 
the number of lines with similar or larger intensities was so large that it would be 
difficult to identify unblended lines unambiguously. In addition, it is important to note 
that $\Delta K_a = 1$ transitions carry less information on the purely $K$-dependent 
parameters than transitions with $\Delta K_a = 2$ at similar values of $K_a$.


\begin{table}
\begin{center}
\caption{Spectroscopic parameters (MHz) from a minimum parameter set, a full parameter set, 
         and initial parameters along with details on the fits of the mono and doubly deuterated 
         isotopologs of cyclopropenone.}
\label{tab-deuterated-isos}
\renewcommand{\arraystretch}{1.10}
\begin{tabular}[t]{lr@{}lr@{}lr@{}l}
\hline \hline
Parameter & \multicolumn{2}{c}{Minimum} & \multicolumn{2}{c}{Full} & \multicolumn{2}{c}{Initial} \\
\hline
\multicolumn{7}{l}{HD isotopolog}                                        \\
\hline
$A$                  &  28049&.092~(3)  &  28049&.090~(4)  &  28035&.946 \\
$B$                  &   7510&.3156~(1) &   7510&.3156~(1) &   7509&.008 \\
$C$                  &   5916&.1823~(1) &   5916&.1823~(1) &   5915&.416 \\
$D_K \times 10^3$    &     46&.16       &     45&.95~(33)  &     46&.16  \\
$D_{JK} \times 10^3$ &     29&.541~(3)  &     29&.541~(3)  &     29&.509 \\
$D_J \times 10^6$    &   1347&.93~(15)  &   1347&.99~(18)  &   1349&.3   \\
$d_1 \times 10^6$    & $-$412&.68~(8)   & $-$412&.71~(9)   & $-$414&.44  \\
$d_2 \times 10^6$    & $-$180&.56~(68)  & $-$180&.54~(72)  & $-$181&.29  \\
$J_{\rm max}$        &     20&~(0)      &     20&~(0)      &       &     \\
$K_{a,{\rm max}}$    &      7&~(0)      &      7&~(0)      &       &     \\
No. trans.           &    130&~(0)      &    130&~(0)      &       &     \\
No. freqs.           &    126&~(0)      &    126&~(0)      &       &     \\
rms                  &      0&.0126     &      0&.0122     &       &     \\
wrms$^a$             &      0&.817      &      0&.815      &       &     \\
\\
\multicolumn{7}{l}{D$_2$ isotopolog}                                     \\
\hline
$A$                  &  24535&.524~(1)  &  24535&.521~(3)  &  24514&.520 \\
$B$                  &   7254&.9417~(1) &   7254&.9422~(1) &   7252&.557 \\
$C$                  &   5591&.7853~(1) &   5591&.7855~(1) &   5590&.549 \\
$D_K \times 10^3$    &     30&.37       &     30&.20~(11)  &     30&.37  \\
$D_{JK} \times 10^3$ &     27&.185      &     27&.189~(1)  &     27&.185 \\
$D_J \times 10^6$    &   1179&.79       &   1180&.35~(13)  &   1179&.8   \\
$d_1 \times 10^6$    & $-$410&.13~(10)  & $-$410&.31~(11)  & $-$412&.33  \\
$d_2 \times 10^6$    & $-$199&.06~(4)   & $-$199&.10~(5)   & $-$200&.75  \\
$J_{\rm max}$        &     24&~(10)     &     24&~(10)     &       &     \\
$K_{a,{\rm max}}$    &      9&~(2)      &      9&~(2)      &       &     \\
No. trans.           &    189&~(10)     &    189&~(10)     &       &     \\
No. freqs.           &    172&~(10)     &    172&~(10)     &       &     \\
rms                  &      0&.0222     &      0&.0214     &       &     \\
wrms$^a$             &      0&.917      &      0&.810      &       &     \\
\hline
\end{tabular}
\end{center}
\tablefoot{
Watson's $S$ reduction was used in the representation $I^r$. Numbers in parentheses of parameters are 
one standard deviation in units of the least significant figures. Numbers in parentheses associated with 
numbers of quantum numbers, transitions, and lines, respectively, refer to data from the previous study 
\citep{c-H2C3O_rot_1973}. Parameters without uncertainties were estimated and kept fixed in the analyses, 
see Sect.~\ref{lab-results}. 
$^{(a)}$ Weighted rms, unitless. 
}
\end{table}


The minimum parameter sets for both deuterated isotopologs were determined as described above. 
Interestingly, we obtained a satisfactory fit for 172 different frequencies of $c$-D$_2$C$_3$O 
after floating only two of the five quartic centrifugal distortion parameters, whereas we required 
four for 126 different frequencies of $c$-HDC$_3$O; this is probably a consequence of the choice 
of the covered transitions. In the fit of $c$-D$_2$C$_3$O with the full parameter set, 
$\Delta D_J$ and $\Delta D_{JK}$ improved the wrms by similar amounts. The parameter values, 
their uncertainties, and additional details on the fits are presented 
in Table~\ref{tab-deuterated-isos}.

\section{Structural parameters of cyclopropenone}
\label{structure}

Several methods exist to derive structural parameters of a molecule from its moments of inertia, 
which are inversely proportional to the rotational parameters. The results depend on the model 
to a varying degree, and some methods do not even yield the same result for different data sets 
in theory. Data for different isotopologs are required for molecules with three or more 
symmetry-inequivalent atoms; it is usually desirable to substitute each symmetry-inequivalent atom once to be able to obtain reliable structural parameters.

The ground-state effective (or $r_0$) structure is the most straightforward model in which 
the structural parameters are fit to the ground-state moments of inertia. The ground-state 
values are usually the ones determined first, and are often the only ones. 
The ground-state moments of inertia contain vibrational contributions from the zero-point 
vibrations, causing the $r_0$ structure to be one of the less meaningful structure models, 
with sometimes relatively large changes between isotopic data sets.


\begin{table*}
\begin{center}
\caption{Ground-state rotational parameters $B_{i,0}$ of cyclopropenone isotopic species, vibrational corrections 
         $\Delta B_{i,{\rm v}}$$^a$, resulting semi-empirical equilibrium rotational parameters $B_{i,e}^{\rm SE}$, 
          and inertia defects $\Delta$.$^b$}
\label{tab_equi-parameters}
\renewcommand{\arraystretch}{1.10}
\begin{tabular}[t]{llr@{}lr@{}lr@{}lcr@{}lr@{}lcr@{}lr@{}l}
\hline \hline
 & & & & \multicolumn{4}{c}{B3LYP} & & \multicolumn{4}{c}{MP2} & & \multicolumn{4}{c}{ae-MP2} \\
\cline{5-8} \cline{10-13} \cline{15-18} 
Species & $B_i$/$\Delta$ & \multicolumn{2}{c}{$B_{i,0}$} & \multicolumn{2}{c}{$\Delta B_{i,{\rm v}}$} & 
\multicolumn{2}{c}{$B_{i,e}^{\rm SE}$} & & \multicolumn{2}{c}{$\Delta B_{i,{\rm v}}$} & 
\multicolumn{2}{c}{$B_{i,e}^{\rm SE}$} & & \multicolumn{2}{c}{$\Delta B_{i,{\rm v}}$} & \multicolumn{2}{c}{$B_{i,e}^{\rm SE}$} \\
\hline
main      & $A$      & 32040&.686  & 167&.049 & 32207&.735   & & 179&.586 & 32220&.272   & & 183&.294 & 32223&.980   \\
          & $B$      &  7825&.001  &  33&.674 &  7858&.675   & &  34&.377 &  7859&.378   & &  34&.577 &  7859&.578   \\
          & $C$      &  6280&.728  &  36&.969 &  6317&.697   & &  37&.689 &  6318&.417   & &  37&.999 &  6318&.727   \\
          & $\Delta$ &     0&.1068 &    &     &  $-$0&.00547 & &    &     &  $-$0&.00273 & &    &     &  $-$0&.00321 \\
$^{13}$C1 & $A$      & 32040&.141  & 167&.513 & 32207&.654   & & 180&.110 & 32220&.251   & & 183&.804 & 32223&.945   \\
          & $B$      &  7816&.553  &  33&.050 &  7849&.603   & &  33&.778 &  7850&.331   & &  33&.969 &  7850&.522   \\
          & $C$      &  6275&.289  &  36&.543 &  6311&.832   & &  37&.281 &  6312&.570   & &  37&.584 &  6312&.873   \\
          & $\Delta$ &     0&.1065 &    &     &  $-$0&.00550 & &    &     &  $-$0&.00276 & &    &     &  $-$0&.00324 \\
$^{13}$C2 & $A$      & 31174&.389  & 161&.272 & 31335&.661   & & 173&.586 & 31347&.975   & & 177&.251 & 31351&.640   \\
          & $B$      &  7709&.476  &  32&.969 &  7742&.445   & &  33&.660 &  7743&.136   & &  33&.863 &  7743&.339   \\
          & $C$      &  6172&.746  &  36&.126 &  6208&.872   & &  36&.842 &  6209&.588   & &  37&.155 &  6209&.901   \\
          & $\Delta$ &     0&.1083 &    &     &  $-$0&.00548 & &    &     &  $-$0&.00270 & &    &     &  $-$0&.00321 \\
$^{18}$O  & $A$      & 32040&.339  & 167&.399 & 32207&.738   & & 180&.001 & 32220&.340   & & 183&.743 & 32224&.082   \\
          & $B$      &  7344&.877  &  31&.126 &  7376&.003   & &  31&.751 &  7376&.628   & &  31&.941 &  7376&.818   \\
          & $C$      &  5967&.532  &  34&.415 &  6001&.947   & &  35&.072 &  6002&.604   & &  35&.363 &  6002&.895   \\
          & $\Delta$ &     0&.1079 &    &     &  $-$0&.00537 & &    &     &  $-$0&.00264 & &    &     &  $-$0&.00314 \\
HD        & $A$      & 28049&.092  & 130&.214 & 28179&.306   & & 137&.620 & 28186&.712   & & 140&.330 & 28189&.422   \\
          & $B$      &  7510&.316  &  31&.968 &  7542&.284   & &  32&.521 &  7542&.837   & &  32&.719 &  7543&.035   \\
          & $C$      &  5916&.182  &  34&.032 &  5950&.214   & &  34&.480 &  5950&.662   & &  34&.764 &  5950&.946   \\
          & $\Delta$ &     0&.1142 &    &     &  $-$0&.00590 & &    &     &  $-$0&.00267 & &    &     &  $-$0&.00325 \\
D$_2$     & $A$      & 24535&.524  &  98&.171 & 24633&.695   & & 102&.770 & 24638&.294   & & 105&.086 & 24640&.610   \\
          & $B$      &  7254&.942  &  31&.074 &  7286&.016   & &  31&.600 &  7286&.542   & &  31&.729 &  7286&.671   \\
          & $C$      &  5591&.785  &  31&.521 &  5623&.306   & &  31&.843 &  5623&.628   & &  32&.081 &  5623&.866   \\
          & $\Delta$ &     0&.1210 &    &     &  $-$0&.00641 & &    &     &  $-$0&.00272 & &    &     &  $-$0&.00337 \\
\hline
\end{tabular}
\end{center}
\tablefoot{
$^{(a)}$ $\Delta B_{i,{\rm v}} = \sum_{j} \alpha _j^{B_i}/2$ calculated by different quantum-chemical means 
         as detailed in Sect.~\ref{qcc}. 
$^{(b)}$ All numbers in units of MHz, except $\Delta$ in units of amu\,{\AA}$^2$. 
} 
\end{table*}


The differences in the moments of inertia caused by the substitution of one or more 
symmetry-inequivalent atoms can be used to determine the Cartesian coordinates of these atoms 
\citep{r_s_1953,Costain_1958}. The resulting substitution (or $r_s$) structure reduces 
the effects of vibrational contributions to the ground-state moments of inertia, albeit to 
a varying degree. The $r_s$ structure is equivalent to $r_{\Delta I}$, the structure that 
is obtained by fitting structural parameters to the differences of the moments of inertia 
\citep{r_I-eps_1991}. This structure is, in turn, equivalent to the $r_{I,\epsilon}$ structure, 
in which vibrational contributions $\epsilon _i$ to the ground-state moments of inertia, 
that is $I_{ii,0} = I_{ii,e} + \epsilon _i$, with $i = a, b, c$, are assumed to be equal 
for different isotopologs of a given molecule. The advantage of taking $\epsilon _i$ into 
account explicitly is that rotational parameters of isotopic species to be studied can be 
estimated much more accurately, in particular for atoms with more than two isotopes or for multiply 
substituted isotopologs \citep{use_r_I-eps_c-PrGeH3_1992,use_r_I-eps_SO2ClF_1994,H2CS_isos_rot_2019}, 
if residuals of known isotopologs can be transferred to the values of the desired isotopolog. 
In the case of H$_2$CS, \citet{H2CS_isos_rot_2019} extrapolated the residuals of H$_2$C$^{32}$S, 
H$_2$C$^{33}$S, and H$_2$C$^{34}$S to H$_2$C$^{36}$S, and those of H$_2$C$^{32}$S, H$_2$C$^{34}$S, 
and H$_2^{13}$CS to H$_2^{13}$C$^{34}$S.


\begin{table*}
\begin{center}
\caption{Quantum-chemical and experimental bond lengths (pm) and bond angles (deg) of cyclopropenone.}
\label{struct-parameters}
\renewcommand{\arraystretch}{1.10}
\begin{tabular}[t]{lr@{}lr@{}lr@{}lr@{}lr@{}lr@{}lr@{}l}
\hline \hline
Method$^a$ & \multicolumn{2}{c}{$r$(CO)} & \multicolumn{2}{c}{$r$(C$-$C)} & \multicolumn{2}{c}{$r$(C=C)$^b$} 
& \multicolumn{2}{c}{$r$(CH)} & \multicolumn{2}{c}{$\angle$(OCC)} & \multicolumn{2}{c}{$\angle$(C3C1C2)$^b$} 
& \multicolumn{2}{c}{$\angle$(C1C2H)} \\
\hline
B3LYP/3a                   & 120&.157     & 142&.672     & 133&.964     & 108&.054     & 151&.999     & 56&.001     & 153&.841     \\
B3LYP/4a                   & 120&.000     & 142&.626     & 133&.894     & 108&.005     & 152&.005     & 55&.989     & 153&.852     \\
B3LYP/4aC                  & 119&.989     & 142&.613     & 133&.887     & 108&.018     & 152&.004     & 55&.991     & 153&.856     \\
MP2/3a                     & 120&.681     & 143&.399     & 135&.270     & 107&.970     & 151&.858     & 56&.284     & 154&.166     \\
MP2/4a                     & 120&.388     & 143&.046     & 134&.907     & 107&.850     & 151&.865     & 56&.270     & 154&.150     \\
MP2/5a                     & 120&.313     & 142&.962     & 134&.815     & 107&.817     & 151&.868     & 56&.264     & 154&.142     \\
ae-MP2/3aC                 & 120&.395     & 142&.885     & 134&.752     & 107&.822     & 151&.866     & 56&.268     & 154&.148     \\
ae-MP2/4aC                 & 120&.140     & 142&.612     & 134&.496     & 107&.689     & 151&.865     & 56&.270     & 154&.139     \\
ae-MP2/5aC                 & 120&.078     & 142&.531     & 134&.408     & 107&.655     & 151&.868     & 56&.264     & 154&.132     \\
MP3/3a                     & 119&.628     & 142&.810     & 134&.383     & 107&.698     & 151&.934     & 56&.133     & 153&.478     \\
ae-MP3/3aC                 & 119&.355     & 142&.333     & 133&.908     & 107&.556     & 151&.940     & 56&.121     & 153&.460     \\
CCSD(T)/3a                 & 120&.684     & 143&.735     & 135&.308     & 108&.155     & 151&.921     & 56&.158     & 153&.739     \\
CCSD(T)/4a                 & 120&.346     & 143&.347     & 134&.912     & 108&.030     & 151&.928     & 56&.144     & 153&.731     \\
ae-CCSD(T)/3aC             & 120&.404     & 143&.267     & 134&.832     & 108&.010     & 151&.929     & 56&.142     & 153&.724     \\
$r_s$$^c$                  & 121&.2~(2)   & 141&.2~(3)   & 130&.2~(3)   & 109&.7~(3)   &    &         & 54&.9~(7)   & 152&.5~(5)   \\
$r_{I,\epsilon}$(old)$^d$  & 120&.29~(36) & 143&.18~(46) & 134&.90~(85) & 107&.82~(31) & 151&.89~(22) & 56&.21~(45) & 153&.91~(41) \\
$r_{I,\epsilon}$(old)$^e$  & 120&.47~(30) & 142&.88~(38) & 134&.84~(38) & 107&.88~(16) & 151&.84~(12) & 56&.31~(23) & 154&.02~(22) \\
$r_{I,\epsilon}$(new)$^f$  & 120&.44~(28) & 142&.91~(35) & 134&.73~(14) & 107&.90~(9)  & 151&.88~(8)  & 56&.24~(16) & 153&.95~(14) \\
$r_e^{\rm SE}$(B3LYP)$^g$  & 120&.04~(21) & 142&.91~(17) & 134&.532~(6) & 107&.77~(4)  & 151&.92~(4)  & 56&.16~(7)  & 153&.74~(6)  \\
$r_e^{\rm SE}$(MP2)$^g$    & 120&.05~(10) & 142&.88~(8)  & 134&.480~(4) & 107&.84~(2)  & 151&.93~(2)  & 56&.15~(3)  & 153&.73~(3)  \\
$r_e^{\rm SE}$(ae-MP2)$^g$ & 120&.05~(12) & 142&.87~(10) & 134&.468~(4) & 107&.85~(2)  & 151&.93~(2)  & 56&.14~(4)  & 153&.73~(3)  \\
\hline
\end{tabular}
\end{center}
\tablefoot{
All values from this work unless indicated otherwise. Numbers in parentheses are one 
standard deviation in units of the least significant figures.
$^{(a)}$ Quantum-chemical calculations as detailed in Sect.~\ref{qcc}; structural parameter 
         determinations as described in Sect.~\ref{structure}. 
$^{(b)}$ Derived parameter in present calculations. 
$^{(c)}$ \citet{c-H2C3O_rot_1973}; structural parameters derived from substitution coordinates; 
         $\angle$(C1C2H) calculated from $\angle$(C2C3H) and $\angle$(C3C1C2).
$^{(d)}$ This work; rotational parameters from \citet{c-H2C3O_rot_1973}, 
         but $A$($^{13}$C2) omitted; see also Sects.~\ref{structure}.
$^{(e)}$ This work; rotational parameter derived as described in Sect.~\ref{lab-results} using data 
         from \citet{c-H2C3O_rot_1973} and \citet{c-H2C3O_rot_1990}. 
$^{(f)}$ This work; rotational parameters from this work. 
$^{(g)}$ Quantum-chemical method in parentheses as detailed in Sect.~\ref{structure}; 
         see also Table~\ref{tab_equi-parameters}.
}
\end{table*}

The equilibrium (or $r_e$) structure of a molecule in the potential minimum is the most meaningful structure. 
However, it is also the most elusive structure because data of more than one isotopolog are needed 
for molecules with three or more symmetry-inequivalent atoms, and for each isotopolog we require the 
knowledge of several vibration-rotation parameters according to

\begin{equation}
\label{equi-B}
B_e = B_0 + \frac{1}{2}\sum_{j} \alpha _j^B - \frac{1}{4}\sum_{j \le k} \gamma _{jk}^B - ...
\end{equation}

\noindent
in order to evaluate the equilibrium parameter $B_e$ from the ground-state value $B_0$. Here, 
the $\alpha _j^B$ are first-order vibrational corrections, the $\gamma _{jk}^B$ are second-order 
vibrational corrections, and so on. Equivalent formulations hold for $A_e$ and $C_e$. 
An attractive and lately very common approach is to calculate $\sum_{j} \alpha _j^B/2$ by 
quantum-chemical means to derive semi-empirical equilibrium rotational parameters $B_{i,e}$ 
from the experimental ground-state values \citep{r_e_emp_1998}. Second and higher order 
vibrational contributions are neglected. Numerous quantum-chemical programs are available 
to carry out such calculations; examples are mentioned in Sect.~\ref{qcc}.

We used B3LYP, MP2, and ae-MP2 calculations with a triple zeta basis set to evaluate 
the first-order vibrational corrections for isotopologs of cyclopropenone which are 
summarized in Table~\ref{tab_equi-parameters} together with ground-state values and 
the resulting semi-empirical equilibrium values. The inertia defect $\Delta = 
I_{cc} - I_{bb} - I_{aa}$ is also given for each set of rotational parameters.

We employed the RU111J program \citep{structure_rev_1995} to derive semi-empirical equilibrium 
structural parameters $r_e^{\rm SE}$ for each of the three quantum-chemical calculations. 
The results are given in Table~\ref{struct-parameters} together with structural parameters 
purely from quantum-chemical calculations and with the $r_s$ parameters from \citet{c-H2C3O_rot_1973}. 
The differences between the $r_s$ parameters and our $r_e^{\rm SE}$ parameters turned out 
to be larger than expected and are discussed in detail in Sect.~\ref{disc_struct}. 
We calculated subsequently $r_{I,\epsilon}$ structures, which are equivalent to $r_s$ 
structures as mentioned above, in order to evaluate the dependence of the resulting 
parameters on diverse input data. The first set of input data were our present rotational 
data from six isotopic species as summarized in Table~\ref{tab_equi-parameters}. 
The second set employed our starting values as described in Sect.~\ref{lab-results}. 
The third set were the rotational parameters from \citet{c-H2C3O_rot_1973}, but without 
$A$($^{13}$C2), as it deviated by 61~MHz from the value of our preferred fit. The results of 
these structure fits are also given (in reverse order) in Table~\ref{struct-parameters}.

\section{Discussion}
\label{discussion}
\subsection{Spectroscopic parameters}
\label{disc_parameters}


As can be seen in Table~\ref{tab-main-species}, the $J$ and $K_a$ quantum number ranges and 
also the number of different transition frequencies of the main isotopolog of cyclopropenone 
have been greatly increased in the course of the present investigation with respect to the 
previous study \citep{c-H2C3O_rot_1990}. The previous spectroscopic parameters agree well 
with ours in the A reduction taking into account the uncertainties in that work. 
The B3LYP/3a quantum-chemically calculated ground-state rotational and equilibrium quartic 
centrifugal distortion parameters agree quite well with our experimental values 
in both reductions. The comparison of the sextic centrifugal distortion parameters 
is more mixed as far as experimental values are available; trial fits showed that 
the experimental values are only slightly affected by the absence of three and one 
sextic parameter in the S and A reduction, respectively. This shows that the deviations 
seen in particular in $H_{KJ}$ and the corresponding $\Phi_{KJ}$ are not mainly caused 
by the omission of the three or one parameter. Moreover, a trial fit with $h_1$ in 
the fit yielded a value of $19 \pm 10~\mu$Hz, not compatible with the calculated 
96~$\mu$Hz.

Our final fit in the S reduction has two parameters less than the one in the A reduction 
at the expense of a slightly larger wrms. We consider this improvement to be sufficient 
to favor the S reduction. Additional reasons for preferring the S reduction are that it 
is more versatile than the A reduction as the latter can exhibit convergence problems 
for molecules close to the prolate or oblate symmetric limit. 
Moreover, \citet{watson_distortion_1977} recommended to use the S reduction because 
such fits yield smaller correlation coefficients. This conclusion was recently reemphasized 
in an extensive study of the rotational spectra of SO$^{18}$O and S$^{18}$O$_2$ in their 
lowest two and three vibrational states, respectively, with detailed analyses of the fits 
\citep{18O-SO2_rot_AuS-red_2020}.

Our spectroscopic parameters for the $^{13}$C1, $^{13}$C2, and $^{18}$O isotopologs in 
Table~\ref{tab-rare-isos} agree very well with our initial values both for the minimum 
and maximum parameter sets. The agreement is good in the cases of the deuterated isotopologs, 
as can be seen in Table~\ref{tab-deuterated-isos}. The parameter $D_K$ from the full parameter 
sets of both deuterated isotopologs agrees very well with the initial values, which may 
indicate that the small changes in $D_K$ seen for the isotopomers containing one $^{13}$C 
may nevertheless be too large.

The initial spectroscopic parameters of all minor isotopic species were derived from 
quantum-chemically calculated ones scaled with the respective ratios between experimental 
and calculated values of the main species as described in slightly more detail at the beginning 
of Sect.~\ref{lab-results}. Such scaling was used, for example, to evaluate some sextic 
centrifugal distortion parameters of two minor isotopologs of TiO$_2$ \citep{TiO2_rot_2011}. 
\citet{scaling_H2S-36_2014} demonstrated in a study of the rotational spectrum of H$_2^{36}$S 
that such scaling works very well even for heavy-atom substitutions of relatively light molecules 
such as H$_2$S. Higher level quantum-chemical calculations, as employed by \citet{scaling_H2S-36_2014}, 
may lead to better starting values than the lower level calculations used here. However, higher 
level calculations are computationally more demanding, in particular for somewhat larger molecules. 
Moreover, \citet{H2CO_ai_2018} showed that considerable computational effort is necessary 
to obtain good-quality, systematically converged results even for the fairly small molecule 
formaldehyde, H$_2$CO. The quality of the scaling of centrifugal distortion parameters may be 
limited in addition by the lack of vibrational corrections. Quartic and sextic equilibrium 
centrifugal distortion parameters can be calculated with several quantum-chemical programs, 
but the derivation of vibrational corrections is, to the best of our knowledge, not possible 
with any publicly available program.

Scaling of quantum-chemically derived distortion parameters should be better than scaling 
by appropriate powers of the ratios of $2A - B - C$, $B + C$, and $B - C$, as done, 
for example, in the recent case of isotopic H$_2$CS \citep{H2CS_isos_rot_2019}. 
This type of scaling  usually works quite well for heavy-atom substitutions, but less so 
for H to D substitutions, in particular for molecules with few atoms. Fixing parameters 
of minor isotopologs, which cannot be determined sufficiently well, to values from 
the main isotopic species usually works less well, but is in most cases better than 
fixing such parameters to zero.


\begin{table*}
\begin{center}
\caption{Equilibrium bond lengths (pm) and selected bond angles (deg) of cyclopropenone in 
         comparison to related molecules.}
\label{struct-comparison}
\renewcommand{\arraystretch}{1.10}
\begin{tabular}[t]{lr@{}lr@{}lr@{}lr@{}lr@{}lr@{}l}
\hline \hline
Molecule & \multicolumn{2}{c}{$r$(CO)} & \multicolumn{2}{c}{$r$(C$-$C)} & \multicolumn{2}{c}{$r$(C=C)} 
& \multicolumn{2}{c}{$r$(CH)} & \multicolumn{2}{c}{$\angle$(C3C1C2)} 
& \multicolumn{2}{c}{$\angle$(C1C2H)} \\
\hline
$c$-H$_2$C$_3$O$^a$ & 120&.05~(12)  & 142&.87~(10) & 134&.468~(4)   & 107&.85~(2)   & 56&.14~(4) & 153&.73~(3) \\
$c$-C$_3$H$_2$$^b$  &    &          & 141&.72~(1)  & 132&.19        & 107&.51~(2)   & 55&.60~(1) & 147&.78~(3) \\
H$_2$CO$^c$         & 120&.461~(19) &    &         &    &           & 110&.046~(16) &   &        &    &        \\
C$_2$H$_6$$^d$      &    &          & 152&.2~(2)   &    &           & 108&.9~(1)    &   &        &    &        \\
$c$-C$_6$H$_6$$^e$  &    &          &\multicolumn{4}{c}{139.14~(10)}& 108&.02~(20)  &   &        &    &        \\
C$_2$H$_4$$^f$      &    &          &    &         & 133&.07~(3)    & 108&.09~(3)   &   &        &    &        \\
\hline
\end{tabular}
\end{center}
\tablefoot{
Numbers in parentheses are one standard deviation in units of the least significant figures. 
If no uncertainties are given, none were published. 
$^{(a)}$ This work, $r_e^{\rm SE}$ with ae-MP2 corrections. 
$^{(b)}$ \citet{c-C3H2_rot_struct_2012}; $r_e^{\rm SE}$, ae-CCSD(T)/4C values; $r$(C=C) calculated 
         from $r$(C$-$C) and $\angle$(C3C1C2).
$^{(c)}$ \citet{H2CO_geo_2006}; experimental $r_e$ values. 
$^{(d)}$ \citet{ethane_r_1990}; $r_m^{\rho}$ values according to the second set of data; see 
         \citet{r-rho-m_1986} for the definition of this structure model. 
$^{(e)}$ \citet{C6H6_struct_2000}; $r_e^{\rm SE}$ at the CCSD(T)/4 level. 
$^{(f)}$ \citet{C2H4_r_e_1996}; $r_e$ values from CCSD(T) calculations extrapolated to infinitely large 
         basis set with ae corrections and empirical adjustments. 
}
\end{table*}


We may ask ourselves what the advantages and disadvantages of our minimum parameter 
fits are compared to our maximum parameter fits. Increasing the parameter set beyond the optimum 
increases the uncertainties, usually the correlation among the parameters, and  frequently affects 
the parameter values outside the initial uncertainties. Therefore, transition 
frequencies calculated with fewer parameters are often better in cases of interpolation 
or modest extrapolation compared to values from a larger parameter set. On the other hand, 
calculated uncertainties quickly become too optimistic upon extrapolation for a smaller 
parameter set. Extrapolation in $J$ from our present data should be reasonable to two 
times the upper frequency limit for the strong $R$-branch transition which is $\sim$1.0~THz 
for the main isotopolog and 450 to 500~GHz for the minor isotopologs. Extrapolation in $K_a$ 
is much more limited.

The ground-state inertia defects in Table~\ref{tab_equi-parameters} are relatively small 
and positive, as is common for planar molecules with small out-of-plane vibrational 
amplitudes. The semi-empirical equilibrium inertia defects obtained from the ground-state 
rotational constants after subtracting the first-order vibrational corrections are 
much smaller in magnitude and negative, which is also very common.

\subsection{Structural parameters}
\label{disc_struct}

Our semi-empirical equilibrium structural parameters in Table~\ref{struct-parameters} are 
largely very similar among the three different sets. The CO bond length has fairly large 
uncertainties, but displays almost no scatter; the C=C bond shows, in contrast, more 
pronounced, though still modest variations, which are only a few times the combined 
uncertainties for the parameters employing B3LYP and ae-MP2 corrections. The CH bond varies 
to an approximately equal degree, albeit with larger uncertainties than the C=C bond.

The previously published $r_s$ structure \citep{c-H2C3O_rot_1973} agrees, at best, 
reasonably well with our $r_e^{\rm SE}$ structures; the C=C bond length difference is, 
with more than 4~pm, particularly large. We calculated $r_{I,\epsilon}$ structures that 
are equivalent to the $r_s$ structure \citep{r_I-eps_1991}. Using our ground-state 
rotational data of six isotopologs, as summarized in Table~\ref{tab_equi-parameters}, 
we obtained structural parameters close to our $r_e^{\rm SE}$ values; the largest 
difference of 0.4~pm occurs in the CO bond length and is even within the combined 
uncertainties. The initial parameters of five isotopic species, as derived in 
Sect.~\ref{lab-results}, yielded essentially the same parameters, the main differences 
are larger uncertainties in some parameters, most likely caused by the absence of data 
of the HD isotopolog; even the fit with values from \citet{c-H2C3O_rot_1973} 
yields results quite similar to the other two $r_{I,\epsilon}$ fits. The deviation 
of the $r_s$ structure parameters  therefore appears to be caused by the difference of 61~MHz 
in $A$($^{13}$C2) which is mainly caused by two lines falsely attributed to the $^{13}$C2 
isotopolog; see also Sect.~\ref{lab-results}.

The quantum-chemical calculations in Table~\ref{struct-parameters} show very small changes 
between basis sets and also rather small changes between the different methods. The bond lengths 
display the usual shortening upon increasing basis set size and upon correlation of all 
electrons as far as applicable. The bond lengths differ substantially among the different 
methods. The B3LYP parameters show little change with basis set size and agree quite well 
with our $r_e^{\rm SE}$ values. The largest deviation is seen for the C=C bond length, 
which is too short by about 0.5~pm. The C$-$C bond length is also too short, but 
by smaller amounts, whereas the CO and CH bonds were calculated slightly too long. 
The MP2/5a and ae-MP2/5aC structures agree between well and very well with our 
$r_e^{\rm SE}$ structure, the MP2/5a values are slightly closer for the C$-$C and CH 
bonds, whereas the ae-MP2/5aC are slightly closer for the other two bond lengths. 
Calculations using MP3 are sometimes better than those with MP2, but are usually worse 
than those with CCSD(T). In the case of cyclopropenone, some bond lengths 
are  already too short at the MP3/3a level, and even more so at ae-MP3/3aC level. 
Bond lengths at the CCSD(T)/3a level are rather long. The effects of correlating all 
electrons compared to those with a frozen core are similar to those of MP2 calculations 
employing basis sets of triple zeta quality. 
If we assume the relative differences are the same for basis sets of quadruple 
zeta quality, we can estimate the following ae-CCSD(T)/4aC bond lengths: 
$r$(CO) $\approx$ 120.10~pm, $r$(C$-$C) $\approx$ 142.95~pm, $r$(C=C) $\approx$ 
134.53~pm, and $r$(CH) $\approx$ 108.00~pm, all very close to our $r_e^{\rm SE}$ 
values; this behavior is common for ae-CCSD(T)/4aC calculations.

Structural parameters of cyclopropenone are compared with those of other molecules 
in Table~\ref{struct-comparison}. Cyclopropenylidene is best suited for comparison 
as it most closely resembles cyclopropenone, but without the O atom. The O atom 
withdraws electron density from the rest of the cyclopropenylidene molecule, causing 
an elongation of all bond lengths; the effect is smallest for the distant CH bond. 
Formaldehyde is probably the prototypical molecule with a CO double bond. 
Its CO bond is only slightly longer than that of $c$-H$_2$C$_3$O. 
The C=C bond lengths of $c$-H$_2$C$_3$O and $c$-C$_3$H$_2$ are essentially equal 
to the one in C$_2$H$_4$, and are thus typical CC double bonds. 
The C$-$C bonds of both cyclic molecules are nevertheless much shorter than that 
of ethane, whose C$-$C bond can probably be viewed as a prototypical CC 
single bond. We note that these bonds are only slightly longer than the aromatic 
CC bonds in benzene. Finally, the CH bond lengths in $c$-H$_2$C$_3$O and 
$c$-C$_3$H$_2$ are substantially shorter than in C$_2$H$_4$, but much longer 
than $\sim$106.2~pm in acetylene \citep{C2H2_re-SE_2011,C2H2_re-exp_2016}. 
These characteristics highlight the aromatic character of cyclopropenone 
\citep{c-H2C3O_aromatic_1983,aromatic_three-rings_ai_2011} and 
cyclopropenylidene from a structural point of view.

\section{Conclusion and outlook}
\label{conclusion}

Our data for five minor isotopic species of cyclopropenone are sufficient for astronomical 
searches in cold molecular clouds even without any extrapolation, as the Boltzmann peak 
at 10~K is at $\sim$95~GHz. Extrapolation of the data in $J$ should be reasonable up to 
450 or 500~GHz. Such extrapolations should be possible up to 1~THz for the main isotopic 
species, which is sufficient even for rotational temperatures around 300~K. However, we 
assume that rotational temperatures of $c$-H$_2$C$_3$O in warmer sources will be closer 
to 100~K based on findings for propanal, propenal, and propene in IRAS 16293$-$2422~B 
\citep{propanal_etc_2017,propenal_propene_2020}.

Dedicated searches will probably be necessary to detect minor isotopologs of cyclopropenone 
in space. $c$-HDC$_3$O is the most promising isotopolog to be found in cold molecular clouds. 
One of the most sensitive molecular line surveys of such sources is that by 
\citet{det-HC3O+_TMC-1_2020} between 31.4 and 50.3~GHz carried out with the Yebes 40~m 
radio telescope with additional data from the 3~mm region taken earlier with the IRAM 
30~m dish. \citet{det-HC3O+_TMC-1_2020} report the detection of HC$_3$O$^+$ and 
investigated the chemistry of oxygen-containing molecules in TMC-1. They derived 
a column density of $(4.0 \pm 0.2) \times 10^{11}$~cm$^{-2}$ for $c$\nobreakdash-H$_2$C$_3$O. 
The most constraining transition of $c$-HDC$_3$O in that survey is the $3_{13} - 2_{12}$ 
transition at 37833.721~MHz. The $3\sigma$ upper limit to its intensity in brightness 
temperature is $\sim$2.6~mK (with corrections for beam dilution and beam efficiency); 
the corresponding intensity of the $3_{13} - 2_{12}$ \textit{ortho} transition of 
$c$\nobreakdash-H$_2$C$_3$O is 42~mK, yielding a $3\sigma$ column density ratio of 
$> 16$ for $c$\nobreakdash-H$_2$C$_3$O to $c$-HDC$_3$O (J. Cernicharo, private 
communication to HSPM, Nov./Dec. 2020).
This is less constraining than the corresponding ratios of, for example, $\sim$34 
and $\sim$25 found for $c$\nobreakdash-C$_3$H$_2$ to $c$\nobreakdash-C$_3$HD 
\citep{det-c-C3HD_1986} and $l$\nobreakdash-C$_3$H$_2$ to $l$\nobreakdash-C$_3$HD 
\citep{det-l-HDC3_2016}, respectively, in TMC-1. 
However, there are sources with higher degrees of deuteration that may be 
considered if their column densities of $c$-H$_2$C$_3$O are sufficiently large. 
A search for the cyclopropenone isotopomers with one $^{13}$C may be more promising 
in the envelope of Sagittarius~B2(N) because the $^{12}$C to $^{13}$C ratio in the 
Galactic center is as low as 20 to 30 
\citep{13C-VyCN_2008,RSH_ROH_2016,12C-13C_SgrB2N_2017,det-13CH_2020}.

We determined structural parameters of cyclopropenone which compare very favorably with 
those of cyclopropenylidene as far as comparison is possible. The CO bond length agrees 
almost exactly with that of formaldehyde. Relatively large deviations between our structural 
parameters and the earlier substitution structure were traced to misassignments of two 
transitions to the $^{13}$C2 isotopolog which led to an overly large $A$ rotational parameter.


\begin{acknowledgements}
We are grateful to Jos{\'e} Cernicharo for communication of unpublished results of a 
search for $c$-HDC$_3$O in TMC-1. 
We acknowledge support by the Deutsche Forschungsgemeinschaft via the collaborative 
research center SFB~956 (project ID 184018867) project B3 as well as 
the Ger{\"a}tezentrum SCHL~341/15-1 (``Cologne Center for Terahertz Spectroscopy''). 
We thank the Regionales Rechenzentrum der Universit{\"a}t zu K{\"o}ln (RRZK) for 
providing computing time on the DFG funded High Performance Computing System CHEOPS. 
J.-C. G. is grateful for financial support by the Centre National d'Etudes Spatiales 
(CNES; grant number 4500065585) and by the Programme National Physique et Chimie du 
Milieu Interstellaire (PCMI) of CNRS/INSU with INC/INP co-funded by CEA and CNES. 
Our research benefited from NASA's Astrophysics Data System (ADS).
\end{acknowledgements}


\bibliographystyle{aa} 
\bibliography{c-H2C3O} 

\end{document}